\documentclass[runningheads]{llncs}
\usepackage{graphicx}
\usepackage{amsmath,amsfonts}
\usepackage{algorithmic}
\usepackage{graphicx}
\usepackage{textcomp}
\usepackage{xcolor}
\usepackage{pgfplotstable}
\usepackage{floatrow}
\newfloatcommand{capbtabbox}{table}[][\FBwidth]
\usepackage{dirtytalk}  
\usepackage{pgfplots}
\usepackage[hyphens]{url}
\usepackage[hyphenbreaks]{breakurl}
\usepackage[breaklinks=true]{hyperref}
\usepackage{subcaption}
\usepackage{ifthen}
\usetikzlibrary{shapes.arrows}
\usepackage[utf8]{inputenc}
\usepackage{hyphenat}
\newcommand{\LULESH}[0]{\textsc{Lulesh}}

\begin{document}
\title{Portability and Scalability of OpenMP Offloading on State-of-the-art Accelerators}
\titlerunning{Portability and Scalability of OpenMP Offloading on SOTA Accelerators}
\authorrunning{Y. Fridman, G. Tamir and G. Oren}
\author{Yehonatan Fridman\inst{1,2,3} \and Guy Tamir\inst{4} \and Gal Oren\inst{3,5}}

\institute{Department of Computer Science, Ben-Gurion University of the Negev, Israel \and Department of Physics, Nuclear Research Center – Negev, Israel \and Scientific Computing Center, Nuclear Research Center – Negev, Israel \and Intel Corporation \and Department of Computer Science, Technion – Israel Institute of Technology \\ 
\email{fridyeh@post.bgu.ac.il, guy.tamir@intel.com, galoren@cs.technion.ac.il}}

%
%
%
%
%
\maketitle              
\begin{abstract}
\vspace{-0.5cm}

Over the last decade, most of the increase in computing power has been gained by advances in accelerated many-core architectures, mainly in the form of GPGPUs. While accelerators achieve phenomenal performances in various computing tasks, their utilization requires code adaptations and transformations. Thus, OpenMP, the most common standard for multi-threading in scientific computing applications, introduced offloading capabilities between host (CPUs) and accelerators since v4.0, with increasing support in the successive v4.5, v5.0, v5.1, and the latest v5.2 versions. Recently, two state-of-the-art GPUs -- the Intel Ponte Vecchio Max 1100 and the NVIDIA A100 GPUs -- were released to the market, with the oneAPI and NVHPC compilers for offloading, correspondingly. In this work, we present early performance results of OpenMP offloading capabilities to these devices while specifically analyzing the portability of advanced directives (using SOLLVE's OMPVV test suite) and the scalability of the hardware in representative scientific mini-app (the LULESH benchmark). Our results show that the coverage for version 4.5 is nearly complete in both latest NVHPC and oneAPI tools. However, we observed a lack of support in versions 5.0, 5.1, and 5.2, which is particularly noticeable when using NVHPC. From the performance perspective, we found that the PVC1100 and A100 are relatively comparable on the LULESH benchmark. While the A100 is slightly better due to faster memory bandwidth, the PVC1100 reaches the next problem size ($400^3$) scalably due to the larger memory size.
\newline The results are available at: \textcolor{blue}{\url{https://github.com/Scientific-Computing-Lab-NRCN/Accel-OpenMP-Portability-Scalability}}.
\end{abstract}
%
%
%

    
\section{Introduction}

\subsection{Opportunities and Challenges of GPUs}

Supercomputers provide essential infrastructure for groundbreaking research, simulation modeling, and data analysis in various scientific domains~\cite{wilson1989grand}. High-performance computing and storage are critical for achieving high resolution and fidelity and analyzing large outputs. As the demand for performance enhancement of algorithmic classes and associated scientific applications constantly increases, new hardware architectures and designs have been developed~\cite{liu2012accelerating}. Specifically, General-Purpose GPUs (GPGPUs)~\cite{gpgpus} have emerged to reduce the execution time of dense linear algebra and Fast Fourier Transform (FFT) calculations~\cite{chen2011large} that are common in material science, chemistry, astrophysics, and deep learning applications.

GPUs are highly parallel and can perform thousands of computations simultaneously~\cite{bridges2016understanding}. This makes GPUs particularly useful for scientific applications that require massive parallelism, such as simulations, modeling, and data analysis. However, programming GPUs can be challenging. GPUs have complex architectures that require a deep understanding of the hardware and the underlying software stack~\cite{niemeyer2014recent,pallipuram2012comparative,lee2012early}. This complexity can make it challenging for developers to optimize their code for performance. Furthermore, different vendors have their own specific GPU architectures and programming models, and different compilers may have varying levels of support for those architectures and models~\cite{lee2012early}. For example, NVIDIA's CUDA programming model is specific to NVIDIA GPUs, and the NVIDIA HPC compilers (such as \textit{nvcc}) are specifically designed to work with CUDA.

Additional challenges for programming GPUs are related to memory~\cite{qureshi2020tearing}. GPUs are typically used to accelerate compute-intensive tasks that involve large amounts of data. However, transferring data to and from the GPU can be time-consuming, and developers need to optimize their code to minimize data transfer overhead. Also, GPUs have limited memory compared to traditional CPUs, which can be a bottleneck for certain types of applications~\cite{qureshi2020tearing}. Developers need to carefully manage memory usage to avoid performance issues. In addition, debugging GPU code can be more challenging than debugging CPU code due to the complex interactions between the hardware and software stack~\cite{knobloch2020tools}. Developers need to use specialized tools and techniques to debug GPU code effectively.

\subsection{OpenMP for Heterogenous CPU-GPU Computing}

As stated, in recent decades, computing architecture has shifted towards multi-core and many-core shared memory architectures to meet the increasing demand for improved performance. To accommodate these architectures, computational paradigms have been adapted by introducing shared memory parallelism. 

The OpenMP API~\cite{dagum1998openmp} is the most comprehensive API that implements the shared memory model. It includes a set of compiler directives (\textit{pragmas}), library routines, and environment variables that allow a program to be executed in parallel within a shared memory environment. One of the reasons for the popularity of the OpenMP API is its flexible and straightforward interface that is readable and easily interpreted~\cite{mattson2019openmp}.


Originally, OpenMP was designed for shared-memory parallel architectures such as multicore CPUs~\cite{mattson2019openmp}. However, with the advent of accelerator devices such as GPUs, there has been a growing demand for OpenMP to support offloading computations to these devices~\cite{van2017using}. OpenMP first introduced support for offloading computations to accelerators in its version 4.0, which was released in 2013~\cite{arb2013openmp}. The OpenMP 4.0 specification added a set of directives for offloading computations to accelerator devices, including the \textit{target} directive and the \textit{map} clause. These directives allow OpenMP programs to offload computations to an accelerator device, such as a GPU, while still maintaining a single codebase for both the host CPU and the accelerator.

Initially, the support for offloading computations to GPUs in OpenMP was experimental and dependent on the specific compiler and runtime library being used. However, support for offloading computations to GPUs in OpenMP has become more widespread in recent years, with many compilers and runtime libraries adding support for OpenMP offloading to GPUs~\cite{OpenMP-Offload}.

In 2018, the OpenMP Architecture Review Board (ARB) released the OpenMP 5.0 specification, which included several new features and enhancements for offloading computations to accelerators, including GPUs~\cite{openmp5}. The OpenMP 5.0 specification includes new directives for managing data transfers between the host and the accelerator, as well as new constructs for managing task dependencies in offloaded computations. The current version is 5.2, released in November 2021~\cite{openmp5.2}.

Today, many widely-used compilers and runtime libraries, including GNU Compiler Collection (GCC)~\cite{GNU-openmp}, Intel oneAPI~\cite{oneapi-gpu}, NVIDIA HPC SDK~\cite{nvidia-openmp}, and Clang~\cite{clang-openmp}, provide support for OpenMP offloading to GPUs. In addition, many GPU vendors, including NVIDIA, Intel, and AMD, have developed tools and libraries that integrate with OpenMP to provide optimized support for offloading computations to their GPUs~\cite{mehta2021evaluating}. Other hardware enhancements also improve efficiency, such as the introduction of high bandwidth memory (HBM)~\cite{jun2017hbm} and DDR5 for memory-bound kernels.

\subsection{Contribution}

The support of various compilers in the latest OpenMP specifications is crucial to utilize GPUs in many systems effectively. Moreover, the hardware design also plays a critical role in the success of the kernels' scalability.
In this work, we present a comprehensive portability and performance scalability evaluation of OpenMP offloading directives on two state-of-the-art comparable accelerators (Intel PVC1100 and NVIDIA A100, described at Section \ref{section:accelerators}) using the latest oneAPI~\cite{oneapi-compiler-git} and NVHPC compilers~\cite{nvhpc-compiler}, respectively. In order to test the portability of OpenMP for said accelerators, we use the state-of-the-art SOLLVE OMPVV test suite (\ref{section:sollve}), while for the scalability testing, we use the LULESH benchmark (\ref{sec:lulesh}) varient~\cite{openmpgpululesh,lulesh-amd}, which supports OpenMP offloading. 

The findings of our study demonstrate that the latest oneAPI and NVHPC compilers provide support for most of the offloading directives included in OpenMP v4.5. However, support for these directives in OpenMP versions 5.0, 5.1 and 5.2 is currently insufficient, specifically in the NVHPC compiler, while their importance is immense. In terms of performance, we noticed that both PVC1100 and A100 accelerators produced similar results for the LULESH benchmark but with a slight advantage for A100, which was around 34\% better. Nevertheless, the PVC100 was capable of running a larger problem ($400^3$), while the A100 experienced memory exhaustion. We also suggest further specific needed support, which can dramatically affect said results.

\section{State-of-the-art Examined HPC Accelerators}\label{section:accelerators}

\subsection{Intel Data Center Max 1100 GPU (Ponte Vecchio, PVC)}\label{section:pvc}

Intel Data Center Max GPU (also known as Ponte Vecchio, in short PVC)~\cite{intel-pvc-web} is a new high-performance computing GPU architecture that is being developed specifically for exascale computing and artificial intelligence workloads. One of the main advantages of Ponte Vecchio is its use of Intel's advanced packaging technology, known as Foveros~\cite{foveros}. This allows multiple chips to be stacked on top of each other in a 3D configuration, which can lead to significant improvements in performance and power efficiency. Another key feature of Ponte Vecchio is its use of Intel's Xe architecture~\cite{intel-xe}, which is designed specifically for high-performance computing and AI workloads. Xe includes specialized hardware for tasks like matrix multiplication, which are common in AI training and inference. Ponte Vecchio is also designed to work with Intel's oneAPI software stack~\cite{oneapi-gpu2}, which is an open, unified programming model that allows developers to write code that can run on a wide variety of processors, including CPUs, GPUs, and FPGAs. This can make it easier for developers to optimize their code for Ponte Vecchio and other Intel processors. While the PVC is available in three versions (1100, 1350, and 1550), we consider the 1100 version relatively comparable to NVIDIA's A100 GPU~\cite{intel-pvc}.

\subsection{NVIDIA A100 GPU}\label{section:a100}

The NVIDIA A100 GPU~\cite{a100} is a flagship GPU from NVIDIA designed for high-performance computing, data analytics, and artificial intelligence workloads. It features a 3rd-Generation Tensor Core that provides significant improvements in deep learning performance, as well as an enhanced L2 cache and HBM2 DRAM for faster memory access. The A100 GPU also includes a new asynchronous data movement and programming model, which improves efficiency, and 3rd-generation NVIDIA NVLink I/O, which enables fast and efficient interconnectivity between multiple GPUs \cite{choquette2021nvidia}. The A100 GPU is optimized to function with the NVHPC~\cite{nvhpc-compiler} compilers and libraries suite, which is specifically designed for NVIDIA GPU architectures. Overall, the NVIDIA A100 GPU is a highly powerful and versatile solution for demanding computing workloads across various industries and applications. 

\section{OpenMP Offloading Port' \& Scal' Benchmarks}\label{section:benchmarks}

\subsection{SOLLVE OpenMP Validation and Verification (OMPVV)}\label{section:sollve}

The OpenMP language is continuously developing with the release of each new specification, resulting in the need to validate and verify the new features implemented by different vendors. The latest versions of OpenMP, 5.0, 5.1, and 5.2, have introduced many new target offload and host-based features to the programming model, indicating the growing maturity of OpenMP and the increasing number of compiler and hardware vendors that support it~\cite{OpenMP-Offload}. The SOLLVE Validation and Verification test suite~\cite{sollve-web} was built to provide open-source vendor-agnostic feature tests for the latest OpenMP specifications with a focus on features of interest to applications. The SOLLVE OMPVV project~\cite{sollve-web} focuses on evaluating the implementation progress and conformity of various compiler vendors, such as Intel, NVIDIA, Clang/LLVM, IBM, GNU, and Cray, for the 4.5, 5.0, 5.1, and 5.2 versions of the specification. The effort of fault-finding in these implementations is particularly valuable for application developers using new OpenMP features to speed up their scientific codes. Huber et al. \cite{huber2022ecp} provide insights into the current implementation status of different vendors, the progression of specific compilers' OpenMP support over time, and examples of how the test suite has influenced discussions regarding the correct interpretation of the OpenMP specification. However, this work does not show results for the state-of-the-art GPUs. 

\subsection{OpenMP Offloading LULESH}\label{sec:lulesh}

\LULESH{} is a well-known mini-application that simplifies the simulation of the Sedov-Taylor problem~\cite{sedovtaylor} using unstructured Lagrangian rezoning to solve hydrodynamic equations (Navier-Stokes) explicitly on an unstructured grid. This makes it latency and bandwidth-bound~\cite{wen2018profdp}. \LULESH{} supports various parallelization schemes and programming paradigms such as MPI, OpenMP, CUDA, OpenACC, and OpenCL~\cite{luleshwebpage}. It was developed as part of one of the five challenge problems defined by the DARPA UHPC program~\cite{darpa,luleshwebpage} and aims to simulate a small portion of various multi-physics applications (specifically, ALE3D~\cite{noble2017ale3d}) that consume up to 30\% of the computing resources of the DoD and DoE~\cite{openmpgpululesh,luleshtuningale3d,hornung2011hydrodynamics}. Due to its popularity, \LULESH{} is widely used for testing new hardware, applying optimizations, parallelization schemes, APIs, and more~\cite{scalingthesummit,openmpgpululesh}. For instance, Laney et al.~\cite{laney2013assessing} used \LULESH{} and several other benchmarks to assess the effects of data compression on performance, as they are representative of real scientific applications. Bercea et al.~\cite{openmpgpululesh} evaluated the effectiveness of the OpenMP offloading capabilities on NVIDIA Kepler GPUs K40m (2015), and it is the latest known version of its kind, using OpenMP version 4.0~\cite{openmpgpululesh}.



\section{Results and Analysis}
\subsection{Settings and Hardware}

Both NVIDIA A100 and Intel Ponte Vecchio MAX are designed for high-performance computing and AI workloads. The A100 is based on NVIDIA's Ampere architecture, while the Ponte Vecchio MAX is based on Intel's Xe architecture~\cite{pvc-a100}. These two accelerators are considered comparable (in the PVC1100 version), and their specifications are listed in Table~\ref{gpu-specs}. In this work, we compare OpenMP portability and performance scalability on two systems, as Table \ref{systems} presents.
To compile OpenMP pragmas we use the latest Intel oneAPI DPC++/C++ Compiler (2023) for system \#1 and the latest NVHPC 23.3 for system \#2. To compile \LULESH{} and SOLLVE OMPVV tests with OpenMP offloading, we use supportive flags, as presented in Table~\ref{compilation}.

    \begin{table}[H]
    \centering
        \begin{tabular}{|c |c |c |} 
         \hline
          & \textbf{Intel PVC1100} & \textbf{NVIDIA A100} \\  
         \hline
         GPU Architecture & Xe-HPC & NVIDIA Ampere \\          
         \hline
         Memory & 48GB HBM2e & 40GB HBM2e \\          
         \hline
         Memory Bandwidth & 1228.8 GB/s & 1555 GB/s \\          
         \hline
         Compute Cores & 7168 & 6912 \\          
         \hline
        \end{tabular}
        \caption{Intel Data Center GPU Max 1100 (PVC) specifications~\cite{pvc-1100} v.s. NVIDIA A100 Tensor Core PCIe~\cite{a100-spec}.}
        \label{gpu-specs}
\end{table}
\vspace{-1cm}
   \begin{table}[H]
    \centering
        \resizebox{\textwidth}{!}{\begin{tabular}{|c |c |c |c |} 
         \hline
         \textbf{System} &\textbf{ CPU (host)} & \textbf{GPU (device)} & \textbf{Compiler}\\  
         \hline
         \#1 & $\times2$ Intel 4th Gen Xeon & Intel Data Center GPU & oneAPI 2023.1.0  \\ & (Sapphire Rapids) processors~\cite{intel-sr} & Max 1100~\cite{pvc-1100} &  ifx/icpx/icx \\          
         \hline
         \#2 & $\times2$ Intel Xeon Gold 6338 & NVIDIA A100 Tensor & NVHPC 23.3 \\ & processors~\cite{intel-xeon} & Core GPU~\cite{a100} &  nvfortran/nvc++/nvc \\
         \hline
        \end{tabular}}
        \caption{Compared systems in this work and the compilers used in each system.}
        \label{systems}
\end{table}
\vspace{-1cm}
   \begin{table}[H]
    \centering
        \resizebox{\textwidth}{!}{\begin{tabular}{|c |c |} 
         \hline
         \textbf{System} & \textbf{Compilation flags}\\  
         \hline
         \#1 & \texttt{-O3 -qopenmp -fopenmp-targets=spir64 -fiopenmp \color{gray}{-fopenmp-version=\{50,51,52\}}} \\          
         \hline
         \#2 & \texttt{-O3 -mp=gpu -gpu=cc80} \\
         \hline
        \end{tabular}}
        \caption{Compilation flags in each system listed in Table~\ref{systems}.}
        \label{compilation}
\end{table}
\vspace{-1cm}

\subsection{OpenMP Portability Evaluation with SOLLVE OMPVV}
The results of SOLLVE OMPVV are presented in Figures \ref{fig:sollve_4.5}, \ref{fig:sollve_5.0}, \ref{fig:sollve_5.1} and \ref{fig:sollve_5.2} for OpenMP v4.5, v5.0, v5.1 and v5.2 respectively\footnote{The results are fully listed in \url{https://github.com/Scientific-Computing-Lab-NRCN/Accel-OpenMP-Portability-Scalability/tree/main/Sollve_vv}}. The total number of tests is 603 (230 for v4.5, 281 for v5.0, 80 for v5.1, and 12 for v5.2). The tests cover OpenMP pragmas for C, C++, and Fortran programs, although the coverage for C++ and Fortran is lacking. The test outcomes are labeled as "PASS" for those that successfully pass compilation and execution, and "FAIL" for those that fail during compilation or execution. The results show that both oneAPI 2023 and NVHPC 23.3 compilers provide almost complete coverage for v4.5 directives on the corresponding systems. While oneAPI provides relatively good support for OpenMP v5.0 and v5.1 directives (yet still far from completion), the support for v5.0 and v5.1 by NVHPC is lacking. In addition, both compilers provide lacking support for v5.2 directives. Our analysis reveals limited progress in support of NVHPC for OpenMP directives since 2021, as similar results were reported in~\cite{huber2022ecp} for NVHPC in version 21 on Summit~\cite{wells2016announcing}, and have not been improved upon since.
Table~\ref{fig:sollve_examples} lists some inconsistencies in the results between oneAPI 2023 and NVHPC 23.3. However, there are directives with immense importance that are not supported by both compilers. For example, both systems do not support reverse offloading and unified memory capabilities introduced in OpenMP v5.0 and later (Table \ref{fig:sollve_fail} lists noticeable ones). The reverse offloading directive allows for seamless offloading of work to accelerators, while shared and unified memory directives simplify the memory management between the CPU and accelerator. With strong compiler support, these OpenMP directives become accessible and widely used, resulting in improved performance and productivity in parallel computing applications.


\newcounter{groupcount}
\pgfplotsset{
    draw group line/.style n args={5}{
        after end axis/.append code={
            \setcounter{groupcount}{0}
            \pgfplotstableforeachcolumnelement{#1}\of\datatable\as\cell{%
                \def\temp{#2}
                \ifx\temp\cell
                    \ifnum\thegroupcount=0
                        \stepcounter{groupcount}
                        \pgfplotstablegetelem{\pgfplotstablerow}{X}\of\datatable
                        \coordinate [yshift=#4] (startgroup) at (axis cs:\pgfplotsretval,0);
                    \else
                        \pgfplotstablegetelem{\pgfplotstablerow}{X}\of\datatable
                        \coordinate [yshift=#4] (endgroup) at (axis cs:\pgfplotsretval,0);
                    \fi
                \else
                    \ifnum\thegroupcount=1
                        \setcounter{groupcount}{0}
                        \draw [
                            shorten >=-#5,
                            shorten <=-#5
                        ] (startgroup) -- node [anchor=base, yshift=0.5ex] {#3} (endgroup);
                    \fi
                \fi
            }
            \ifnum\thegroupcount=1
                        \setcounter{groupcount}{0}
                        \draw [
                            shorten >=-#5,
                            shorten <=-#5
                        ] (startgroup) -- node [anchor=base, yshift=0.5ex] {#3} (endgroup);
            \fi
        }
    }
}

\begin{figure}[H]
\begin{tikzpicture}
\pgfplotstableread{
X   Gp           Name     fortran  cpp    c
1   oneAPI+PVC   PASS     98       14     109
2   oneAPI+PVC   FAIL     4        0      5   
3   NVHPC+A100     PASS   97       14     112
4   NVHPC+A100     FAIL   5        0      2

}\datatable

\hspace*{-1.1cm}
\begin{axis}[
    reverse legend,
    axis lines*=left, ymajorgrids,
    width=8cm, height=4cm,
    ymin=0,
    ybar stacked,
    ylabel={Number of tests},
    ytick={0,50,100,150,200,250},
    bar width=20pt,
    xtick=data,
    xticklabels from table={\datatable}{Name},
    draw group line={Gp}{oneAPI+PVC}{oneAPI+PVC}{-7ex}{7pt},
    draw group line={Gp}{NVHPC+A100}{NVHPC+A100}{-7ex}{7pt},
    after end axis/.append code={
        \path [anchor=base east, yshift=0.5ex]
            (rel axis cs:0,0) node [yshift=-16ex,xshift=-2ex] {C}
            (rel axis cs:0,0) node [yshift=-13ex,xshift=-2ex] {C++}
            (rel axis cs:0,0) node [yshift=-10ex,xshift=-2ex] {Fortran}
            (rel axis cs:0,0) node [yshift=-7ex,xshift=-2ex] {Compiler+GPU};
    },
    after end axis/.append code={
        \path [anchor=base east, yshift=0.5ex]
            (rel axis cs:0.13,0) node [yshift=-16ex] {109}
            (rel axis cs:0.13,0) node [yshift=-13ex] {14}
            (rel axis cs:0.13,0) node [yshift=-10ex] {98};
    },
    after end axis/.append code={
        \path [anchor=base east, yshift=0.5ex]
            (rel axis cs:0.4,0) node [yshift=-16ex] {5}
            (rel axis cs:0.4,0) node [yshift=-13ex] {0}
            (rel axis cs:0.4,0) node [yshift=-10ex] {4};
    },
      after end axis/.append code={
        \path [anchor=base east, yshift=0.5ex]
            (rel axis cs:0.7,0) node [yshift=-16ex] {112}
            (rel axis cs:0.7,0) node [yshift=-13ex] {14}
            (rel axis cs:0.7,0) node [yshift=-10ex] {97};
    },
    after end axis/.append code={
        \path [anchor=base east, yshift=0.5ex]
            (rel axis cs:0.95,0) node [yshift=-16ex] {2}
            (rel axis cs:0.95,0) node [yshift=-13ex] {0}
            (rel axis cs:0.95,0) node [yshift=-10ex] {5};
    }
]

\addplot table [x=X, y=c] {\datatable}; \addlegendentry{C}
\addplot table [x=X, y=cpp] {\datatable}; \addlegendentry{C++}
\addplot table [x=X, y=fortran] {\datatable}; \addlegendentry{Fortran}

\end{axis}
\draw (0, 0) rectangle (6.5,2.5);
\end{tikzpicture}
\caption{OpenMP 4.5 tests with oneAPI \& NVHPC for PVC1100 \& A100, respectively.}
\label{fig:sollve_4.5}
\end{figure}
\vspace{-1cm}
\begin{figure}[H]
\begin{tikzpicture}
\pgfplotstableread{
X   Gp           Name     fortran  cpp    c
1   oneAPI+PVC   PASS     69       12     123
2   oneAPI+PVC   FAIL     39       1      37   
3   NVHPC+A100     PASS     33       8     42
4   NVHPC+A100     FAIL     75       5     118

}\datatable

\hspace*{-1cm}
\begin{axis}[
    reverse legend,
    axis lines*=left, ymajorgrids,
    width=8cm, height=4cm,
    ymin=0,
    ybar stacked,
    ytick={0,50,100,150,200,250},
    ylabel={Number of tests},
    bar width=20pt,
    xtick=data,
    xticklabels from table={\datatable}{Name},
    draw group line={Gp}{oneAPI+PVC}{oneAPI+PVC}{-7ex}{7pt},
    draw group line={Gp}{NVHPC+A100}{NVHPC+A100}{-7ex}{7pt},
    after end axis/.append code={
        \path [anchor=base east, yshift=0.5ex]
            (rel axis cs:0,0) node [yshift=-16ex,xshift=-2ex] {C}
            (rel axis cs:0,0) node [yshift=-13ex,xshift=-2ex] {C++}
            (rel axis cs:0,0) node [yshift=-10ex,xshift=-2ex] {Fortran}
            (rel axis cs:0,0) node [yshift=-7ex,xshift=-2ex] {Compiler+GPU};
    },
    after end axis/.append code={
        \path [anchor=base east, yshift=0.5ex]
            (rel axis cs:0.13,0) node [yshift=-16ex] {123}
            (rel axis cs:0.13,0) node [yshift=-13ex] {12}
            (rel axis cs:0.13,0) node [yshift=-10ex] {69};
    },
    after end axis/.append code={
        \path [anchor=base east, yshift=0.5ex]
            (rel axis cs:0.4,0) node [yshift=-16ex] {37}
            (rel axis cs:0.4,0) node [yshift=-13ex] {1}
            (rel axis cs:0.4,0) node [yshift=-10ex] {39};
    },
      after end axis/.append code={
        \path [anchor=base east, yshift=0.5ex]
            (rel axis cs:0.7,0) node [yshift=-16ex] {42}
            (rel axis cs:0.7,0) node [yshift=-13ex] {8}
            (rel axis cs:0.7,0) node [yshift=-10ex] {33};
    },
    after end axis/.append code={
        \path [anchor=base east, yshift=0.5ex]
            (rel axis cs:0.95,0) node [yshift=-16ex] {118}
            (rel axis cs:0.95,0) node [yshift=-13ex] {5}
            (rel axis cs:0.95,0) node [yshift=-10ex] {75};
    }
]

\addplot table [x=X, y=c] {\datatable}; 
\addplot table [x=X, y=cpp] {\datatable}; 
\addplot table [x=X, y=fortran] {\datatable}; 

\end{axis}
\draw (0, 0) rectangle (6.5,2.5);
\end{tikzpicture}
\caption{OpenMP 5.0 tests with oneAPI \& NVHPC for PVC1100 \& A100, respectively.}
\label{fig:sollve_5.0}
\end{figure}
\vspace{-1cm}
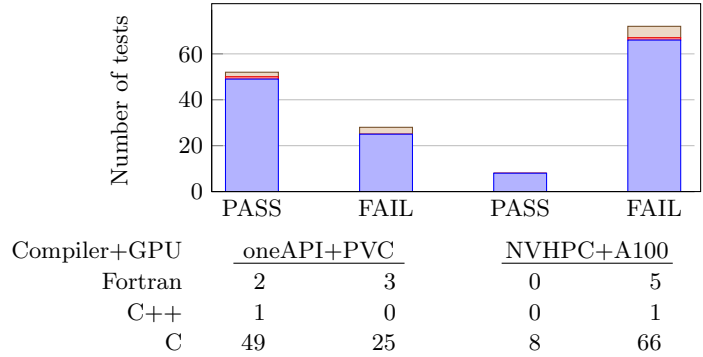
\begin{figure}[H]
\begin{tikzpicture}
\pgfplotstableread{
X   Gp           Name     fortran  cpp    c
1   oneAPI+PVC   PASS     2        1      49
2   oneAPI+PVC   FAIL     3        0      25   
3   NVHPC+A100     PASS     0        0      8
4   NVHPC+A100     FAIL     5        1      66

}\datatable

\hspace*{-1cm}
\begin{axis}[
    reverse legend,
    axis lines*=left, ymajorgrids,
    width=8cm, height=4cm,
    ymin=0,
    ybar stacked,
    ylabel={Number of tests},
    bar width=20pt,
    xtick=data,
    xticklabels from table={\datatable}{Name},
    draw group line={Gp}{oneAPI+PVC}{oneAPI+PVC}{-7ex}{7pt},
    draw group line={Gp}{NVHPC+A100}{NVHPC+A100}{-7ex}{7pt},
    after end axis/.append code={
        \path [anchor=base east, yshift=0.5ex]
            (rel axis cs:0,0) node [yshift=-16ex,xshift=-2ex] {C}
            (rel axis cs:0,0) node [yshift=-13ex,xshift=-2ex] {C++}
            (rel axis cs:0,0) node [yshift=-10ex,xshift=-2ex] {Fortran}
            (rel axis cs:0,0) node [yshift=-7ex,xshift=-2ex] {Compiler+GPU};
    },
    after end axis/.append code={
        \path [anchor=base east, yshift=0.5ex]
            (rel axis cs:0.13,0) node [yshift=-16ex] {49}
            (rel axis cs:0.13,0) node [yshift=-13ex] {1}
            (rel axis cs:0.13,0) node [yshift=-10ex] {2};
    },
    after end axis/.append code={
        \path [anchor=base east, yshift=0.5ex]
            (rel axis cs:0.4,0) node [yshift=-16ex] {25}
            (rel axis cs:0.4,0) node [yshift=-13ex] {0}
            (rel axis cs:0.4,0) node [yshift=-10ex] {3};
    },
      after end axis/.append code={
        \path [anchor=base east, yshift=0.5ex]
            (rel axis cs:0.7,0) node [yshift=-16ex] {8}
            (rel axis cs:0.7,0) node [yshift=-13ex] {0}
            (rel axis cs:0.7,0) node [yshift=-10ex] {0};
    },
    after end axis/.append code={
        \path [anchor=base east, yshift=0.5ex]
            (rel axis cs:0.95,0) node [yshift=-16ex] {66}
            (rel axis cs:0.95,0) node [yshift=-13ex] {1}
            (rel axis cs:0.95,0) node [yshift=-10ex] {5};
    }
]

\addplot table [x=X, y=c] {\datatable}; 
\addplot table [x=X, y=cpp] {\datatable}; 
\addplot table [x=X, y=fortran] {\datatable}; 

\end{axis}
\draw (0, 0) rectangle (6.5,2.5);
\end{tikzpicture}
\caption{OpenMP 5.1 tests with oneAPI \& NVHPC for PVC1100 \& A100, respectively.}
\label{fig:sollve_5.1}
\end{figure}
\vspace{-1cm}

\begin{figure}[H]
\begin{tikzpicture}
\pgfplotstableread{
X   Gp           Name     fortran  cpp    c
1   oneAPI+PVC   PASS     0        0      1
2   oneAPI+PVC   FAIL     4        0      7   
3   NVHPC+A100     PASS     2        0      1
4   NVHPC+A100     FAIL     2        0      7

}\datatable

\hspace*{-1cm}
\begin{axis}[
    reverse legend,
    axis lines*=left, ymajorgrids,
    width=8cm, height=4cm,
    ymin=0,
    ymax=15,
    ybar stacked,
    ylabel={Number of tests},
    bar width=20pt,
    xtick=data,
    xticklabels from table={\datatable}{Name},
    draw group line={Gp}{oneAPI+PVC}{oneAPI+PVC}{-7ex}{7pt},
    draw group line={Gp}{NVHPC+A100}{NVHPC+A100}{-7ex}{7pt},
    after end axis/.append code={
        \path [anchor=base east, yshift=0.5ex]
            (rel axis cs:0,0) node [yshift=-16ex,xshift=-2ex] {C}
            (rel axis cs:0,0) node [yshift=-13ex,xshift=-2ex] {C++}
            (rel axis cs:0,0) node [yshift=-10ex,xshift=-2ex] {Fortran}
            (rel axis cs:0,0) node [yshift=-7ex,xshift=-2ex] {Compiler+GPU};
    },
    after end axis/.append code={
        \path [anchor=base east, yshift=0.5ex]
            (rel axis cs:0.13,0) node [yshift=-16ex] {1}
            (rel axis cs:0.13,0) node [yshift=-13ex] {0}
            (rel axis cs:0.13,0) node [yshift=-10ex] {0};
    },
    after end axis/.append code={
        \path [anchor=base east, yshift=0.5ex]
            (rel axis cs:0.4,0) node [yshift=-16ex] {7}
            (rel axis cs:0.4,0) node [yshift=-13ex] {0}
            (rel axis cs:0.4,0) node [yshift=-10ex] {4};
    },
      after end axis/.append code={
        \path [anchor=base east, yshift=0.5ex]
            (rel axis cs:0.7,0) node [yshift=-16ex] {1}
            (rel axis cs:0.7,0) node [yshift=-13ex] {0}
            (rel axis cs:0.7,0) node [yshift=-10ex] {2};
    },
    after end axis/.append code={
        \path [anchor=base east, yshift=0.5ex]
            (rel axis cs:0.95,0) node [yshift=-16ex] {7}
            (rel axis cs:0.95,0) node [yshift=-13ex] {0}
            (rel axis cs:0.95,0) node [yshift=-10ex] {2};
    }
]

\addplot table [x=X, y=c] {\datatable}; 
\addplot table [x=X, y=cpp] {\datatable}; 
\addplot table [x=X, y=fortran] {\datatable}; 

\end{axis}
\draw (0, 0) rectangle (6.5,2.5);
\end{tikzpicture}
\caption{OpenMP 5.2 tests with oneAPI \& NVHPC for PVC1100 \& A100, respectively.}
\label{fig:sollve_5.2}
\end{figure}
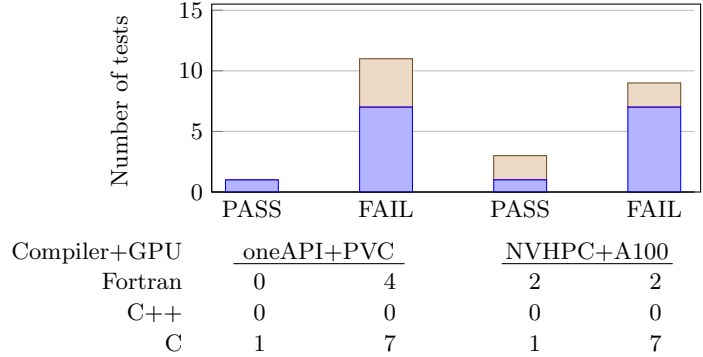
\vspace{-1cm}

\begin{table}[ht]
\resizebox{\textwidth}{!}{\begin{tabular}{ |c|c|c|c| } 
 \hline
 \textbf{Test Name} & \textbf{OMP Version} & \textbf{oneAPI 2023} & \textbf{NVHPC 23.3} \\
\hline
test\_declare\_target\_device\_type\_nohost.c & 5.0  & PASS &  FAIL  \\
\hline
test\_requires\_unified\_shared\_memory & 5.0  & PASS &  FAIL  \\ \_omp\_target\_alloc.c & & &\\
\hline
test\_requires\_unified\_shared\_memory & 5.0  & PASS &  FAIL  \\ \_omp\_target\_alloc\_is\_device\_ptr.c & & &\\
\hline
test\_target\_defaultmap\_present.c & 5.1  & PASS &  FAIL  \\
 \hline
test\_target\_is\_accessible.c & 5.1  & PASS &  FAIL  \\
\hline
 test\_target\_update\_devices.F90  & 4.5  & FAIL &  PASS  \\
\hline
test\_declare\_target\_parallel\_for.F90  & 5.0  & FAIL &  PASS  \\
\hline
test\_requires\_unified\_shared\_memory\_static.F90  & 5.0  & FAIL &  PASS  \\
\hline
test\_metadirective\_target\_device\_kind.c  & 5.1  & FAIL &  PASS  \\
\hline
\end{tabular}}
    \caption{Inconsistencies of SOLLVE OMPVV tests passing and failing on oneAPI+PVC1100 v.s. NVHPC+A100. While some of the failed tests are managed to compile, the others compile successfully but fail to execute.}
    \label{fig:sollve_examples}
\end{table}
\vspace{-1cm}
\begin{table}[ht]
\resizebox{\textwidth}{!}{\begin{tabular}{ |c|c|c|c| } 
 \hline
 \textbf{Test Name} & \textbf{OMP Version} & \textbf{oneAPI 2023} & \textbf{NVHPC 23.3} \\
 \hline
 test\_requires\_reverse\_offload.c & 5.0  & FAIL &  FAIL  \\
 \hline
 test\_requires\_unified\_address.c  & 5.0  & FAIL &  FAIL  \\
 \hline
test\_requires\_unified\_shared\_memory\_static.c  & 5.0  & FAIL &  FAIL  \\
\hline
test\_requires\_unified\_shared\_memory & 5.1  & FAIL & FAIL   \\
\_heap\_is\_device\_ptr.F90 &  & &   \\
 \hline
 test\_requires\_unified\_shared\_memory & 5.1  & FAIL & FAIL   \\
\_stack\_is\_device\_ptr.F90 &  & &   \\
 \hline

\end{tabular}}
    \caption{OpenMP reverse offloading and unified memory\textbackslash addresses directives fail both on oneAPI+PVC1100 and NVHPC+A100. While some of the failed tests are managed to compile, the others compile successfully but fail to execute.}
    \label{fig:sollve_fail}
\end{table}

\subsection{OpenMP Performance Scalability Evaluation with LULESH}
\begin{figure}[H]
\begin{subfigure}[b]{0.55\textwidth}
\centering
\scalebox{.9}{
\begin{tikzpicture}
\begin{axis}[
    xlabel={Problem size},
    xmin=0.05, xmax=420,
    ymin=0, ymax=120,
    xtick={100,200,300,350,400},
    xticklabels={$100^3$,$200^3$,$300^3$,$350^3$,$400^3$},
    legend pos=north west,
    ymajorgrids=true,
    ylabel={Time (sec)},
    grid style=dashed,
    width=6.8cm,
    height=6cm,
    legend style={nodes={scale=0.8, transform shape}},
    extra x tick labels={},
    extra tick style={grid=major,},
]
\addplot[
    color=blue,
    mark=*,
    mark size=2.5pt,
    ]
    coordinates {
    (0,0)(100,6.27)(200,17.21)(300,48.37)(350,76.71)(400,101.07)
    };
\addplot[
    color=orange,
    mark=triangle*,
    mark size=3pt,
    ]
    coordinates {
    (0,0)(100,4.61)(200,14.95)(300,34.11)(350,57.22)
    };
\addplot[
    color=orange,
    mark=star,
    mark size=3pt,
    ]
    coordinates {
    (400,95)
    };
    \legend{PVC, A100}
\end{axis}
\end{tikzpicture}
}
\caption{Total \LULESH{} time for PVC1100 \\ and A100 (lower is better).}
    \label{Fig:LULESH-total}
\end{subfigure}%
\hfill
\pgfplotstableread[row sep=\\,col sep=&]{
    size & speedup \\
    100   & 36.00 \\
    200   & 15.11  \\
    300   & 41.8 \\
    350   & 34.06 \\
    }\mydata
\begin{subfigure}[b]{0.45\textwidth}
\centering
\scalebox{.9}{
\begin{tikzpicture}
    \begin{axis}[
            xlabel={Problem size},
            width=7cm,
            height=5.55cm,
            ymax=60,
            ymin=0,
            xmax=400,
            ybar,
            nodes near coords, 
            ytick={0,10,20,30,40,50,60,70,80}      ,yticklabel=\pgfmathprintnumber{\tick}\,$\%$,
            bar width=18pt,    
            xtick={100,200,300,350},
            xticklabels={$100^3$,$200^3$,$300^3$,$350^3$},
        ]
        \addplot table[x=size,y=speedup,]{\mydata};
    \end{axis}
\end{tikzpicture}
}
\caption{Improvement percentage of A100 over PVC1100 on total \LULESH{} time (see Figure~\ref{Fig:LULESH-total}) (higher is better).}
\label{Fig:LULESH-speedup}
\end{subfigure}%

    \caption{Comparison of \LULESH{} timings on PVC1100 and A100 (for 20 iterations).}
    \label{Fig:LULESH}
\end{figure}
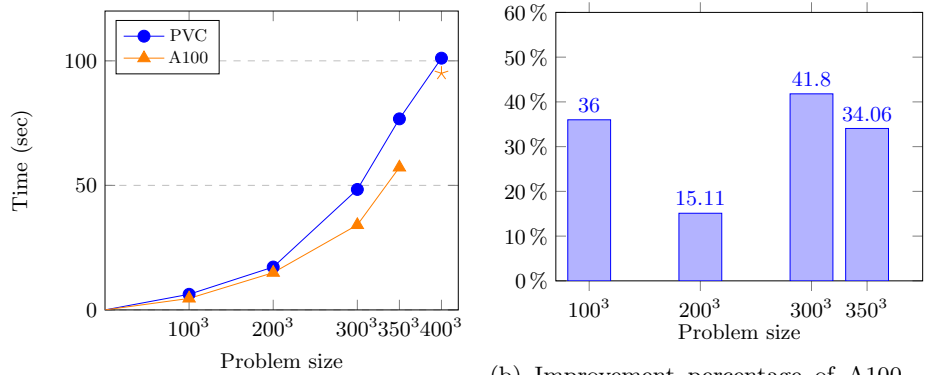

To evaluate the performance scalability of PVC1100 against A100, compilation and execution of \LULESH{} were done with v4.0 OpenMP offloading \cite{lulesh-mp4-git}.
The grid size was scaled from $100^3$ to $400^3$. Each test was run for 20-time steps. We analyze the results by the total \LULESH{} running time\footnote{The \LULESH{} benchmark output results are fully presented in \url{https://github.com/Scientific-Computing-Lab-NRCN/Accel-OpenMP-Portability-Scalability/tree/main/LULESH_Results}} (Figure~\ref{Fig:LULESH-total}). 

In an effort to amplify the outcomes, we made initial endeavors to incorporate support for cutting-edge unified memory\textbackslash addresses capabilities of PVC1100 using the latest directives obtainable on the v5, 5.1, and 5.2 OpenMP specifications. Regrettably, we were unsuccessful in executing the majority of the directives, as shown in Table~\ref{fig:sollve_fail}. Furthermore, the directives that we were able to execute yielded unsatisfactory outcomes or terminated with \textit{\small ZE\_RESULT\_ERROR\_OUT\_OF\_DEVICE\_MEMORY}.

\section{Conclusions and Future Work}
This work comprehensively evaluates the performance scalability and portability of OpenMP offloading on modern compilers and state-of-the-art GPUs. The findings indicate that the latest oneAPI compiler mostly complies with the latest OpenMP specifications, while NVHPC's support is still lacking. In both compilers, support is particularly lacking in the area of unified and shared memory/address directives. Our research shows that the NVIDIA A100  outperforms the Intel PVC1100 (up to \%34 better performance on \LULESH{} with OpenMP 4.0). To further optimize hardware capabilities, we recommend thoroughly exploring integrating the latest OpenMP directives, which can greatly enhance \LULESH{} performance. Specifically, we suggest utilizing \textit{requires unified memory}, and \textit{unified addresses} and \textit{reverse offload} directives to optimize results.

For future work, we propose examining additional proxy apps like XSBench, SNAP, HPGMG, CoMD~\cite{coral,proxy-git}, and ScalSALE~\cite{harel2022scalsale,rusanovsky2019backus} with OpenMP offloading. Due to improvements in compilers support for the newest OpenMP offloading directives that we expect, our goal is to create an enhanced version of LULESH that utilizes these directives to optimize data movements and minimize computational overheads. We further recommend a deeper examination of the portability and scalability of OpenMP offloading in real-world applications using the latest OpenMP directives. Additionally, we propose assessing the scalability of OpenMP in the aforementioned applications across several devices within a single node.


\hfill \\
\noindent
\textbf{Acknowledgments}: This research was supported by Intel Corporation (oneAPI CoE program) and the Lynn and William Frankel Center for Computer Science. Computational support was provided by the NegevHPC project~\cite{negevhpc} and Intel Developer Cloud~\cite{intel-cloud}. The authors want to thank Jay Mahalingam, Omar Toral, Oshana Douglas of Intel, and Israel Hen, Gabi Dadush of NegevHPC for their great help and support.


\bibliographystyle{unsrt} 

\bibliography{bibTex}
\end{document}